\documentclass[useAMS,usenatbib]{mn2e}
\usepackage{lscape}
\usepackage{rotating}
\usepackage{amssymb}
\usepackage{natbib}

\def\msun{\hbox{M$_\odot$}}

\def\t4{\hbox{t$_{\rm 4}$}}

\def\dndt{\hbox{$\frac{d{\rm N}}{dt}$}}

\def\cm3{\hbox{cm$^{-3}$}}

\input psfig.sty
\input epsf.sty
\usepackage{graphicx}
\usepackage{epsfig}
\title[The Age Distribution of Clusters in M83]{The Age Distribution of Stellar Clusters in M83}
\author[Silva-Villa et  al.] {E. Silva-Villa$^{1}$, A. Adamo$^{2}$, N. Bastian$^{3}$, M. Fouesneau$^{4}$, E. Zackrisson$^{5}$\\
$^{1}$ (CRAQ) Universit\'e Laval. 1045, Avenue de la M\'edecine, G1V 0A6 Qu\'ebec, Canada\\
$^{2}$ Max Planck Institut f\"ur Astronomie, K\"onigstuhl 17 D-69117 Heidelberg, Germany\\
$^{3}$ Astrophysics Research Institute, Liverpool John Moores University, 146 Brownlow Hill, Liverpool L3 5RF, UK\\
$^{4}$ Department of Astronomy, University of Washington, 3910 15th Ave NE Seattle, WA 98195-0002, USA\\
$^{5}$ The Oskar Klein Centre, Department of Astronomy, AlbaNova, Stockholm University, SE-106 91 Stockholm, Sweden
}
\date{Accepted. Received; in original form}
\pagerange{\pageref{firstpage}--\pageref{lastpage}}
\pubyear{2013}
\begin{document}
\maketitle
\label{firstpage}
\begin{abstract}
In order to empirically determine the timescale and environmental dependence of stellar cluster disruption, we have undertaken an analysis of the unprecedented multi-pointing (seven), multi-wavelength (U, B, V, H$\alpha$, and I) Hubble Space Telescope imaging survey of the nearby, face-on spiral galaxy M83.  The images are used to locate stellar clusters and stellar associations throughout the galaxy. Estimation of cluster properties (age, mass, and extinction) was done through a comparison of their spectral energy distributions with simple stellar population models. We constructed the largest catalog of stellar clusters and associations in this galaxy to-date, with $\sim1800$ sources with masses above $\sim5000$~\msun and ages younger than $\sim300$ Myr. In the present letter, we focus on the age distribution of the resulting clusters and associations.  In particular, we explicitly test whether the age distributions are related with the ambient environment. Our results are in excellent agreement with previous studies of age distributions in the centre of the galaxy, which gives us confidence to expand out to search for similarities or differences in the other fields which sample different environments. We find that the age distribution of the clusters inside M83 varies strongly as a function of position within the galaxy, indicating a strong correlation with the galactic environment.  If the age distributions are approximated as a power-law of the form $\dndt \propto t^{\zeta}$, we find $\zeta$ values between $0$ and $-0.62$ ($\zeta \sim -0.40$ for the whole galaxy), in good agreement with previous results and theoretical predictions.
\end{abstract}

\section{Introduction}
\label{sec:intro}

The age distribution of stellar cluster populations in galaxies is the result of the combined effects of the cluster formation history and cluster disruption. Studies have shown a tight correlation between the maximum intensity of the star formation history and the age distribution of the cluster population in multiple galaxies (e.g. NGC~7252 - \citet{miller97}; \citet{schweizes98}, \citet{chien10}; M82 - \citet{konstantopoulos09}).  More quiescent galaxies, such as the Milky Way (at least in the solar neighbourhood) have a smoother, rather flat, cluster age distribution \citep[e.g.][]{lamers05,piskunov06}.  

The age distribution of a star cluster population (\dndt) is defined as the number of clusters observed within some linear time interval, and, at present, the role of cluster disruption in setting the shape of the age distribution is still an open topic of debate. If cluster disruption has any dependency with environment, we would expect to see the age distribution changing over different locations, e.g. where there are differences in gas density, or in the tidal field. However, if cluster disruption is independent of environment, the age distributions of any location should be similar.  We chose the galaxy M83 because it presents such differences in environment, allowing us to test these assumptions \citep[see e.g.][for differences in gas densities across M83]{lundgren04}.

Previous studies \citep[e.g.][]{fall05} of the Antennae galaxies (a currently starbursting, merging galaxy) have found that the cluster age distribution is quite steep, with \dndt$ \sim t^{-1}$, and this has been interpreted as evidence for rapid and strong cluster disruption under the assumption that the cluster formation rate/history has been constant over the past few hundred Myr \citep{fall05,whitmore07}. 

However, \citet[][among others]{bastian09} showed that the steepness of the age distribution can be affected by the star formation rate. The age distribution becomes shallower when corrected for an increasing star formation rate, limiting the role of cluster disruption in shaping the age distribution (assuming that both are studied under the same time interval). Also, based on theoretical studies \citep[e.g.][and references there in]{elmegreen10,kruijssen11} it was showed that the ambient environment where a cluster resides can drastically affect the steepness of the age distribution, making it disrupt faster when the surface gas density ($\Sigma_{gas}$) is higher, and live longer when $\Sigma_{gas}$ is low, indicating a strong relation with environment.
 
\citet{chandar06} used the publicly available catalogue of clusters compiled by \citet{hill06} to estimate the age distribution of clusters in the Small Magellanic Cloud (SMC). The authors used a lower mass limit of $10^3$\msun\ and ages between $7\le Log(\tau /yr) \le 9$. They found a steep distribution with \dndt$\sim t^{-0.85}$ in the age interval mentioned.  However, using the same catalogue, \citet{gieles07} and \citet{degrijs08} showed that the steepness in the age distribution was caused by the sample of Hill \& Zaritsky being luminosity-limited. When a mass cut above the completeness was used, the resulting distribution was flat, leading to the conclusion that cluster disruption has not significantly altered the age distribution of SMC clusters.

The Large Magellanic Cloud (LMC) hosts a much larger cluster population than the SMC, and has been the subject of numerous studies. \citet{chandar10} used the \citet{hunter03} catalogue to study the age distribution, and supplement this catalogue with objects that were avoided by Hunter et al., due to nebular emission.  They find a steep age distribution, \dndt$\sim t^{-0.8}$. However, \citet{baumgardt13} and \citet{degrijs13} combine all publicly available catalogues of clusters in the LMC and do not confirm the Chandar et al. findings. Instead they find much flatter results, again suggesting the cluster disruption has not had a strongly effect on the cluster population.

\citet{silvavilla11} studied the age distributions of clusters in five near-by, face-on spiral galaxies, including M83 (NGC~5236) and NGC~1313. Fitting over the age range of 10 to 500~Myr, they found $\dndt \sim t^{-0.25}$ and $t^{-0.65}$ for M83 and NGC~1313, respectively. Their results suggest that the age distribution of star clusters might vary as a function of environment

The Panchromatic Hubble Andromeda Treasury survey \citep[PHAT survey,][]{dalcanton12}, one of the largest studies to-date, has catalogued stellar clusters within M31 \citep[][to date only $\sim25$\% of the survey area has been presented]{johnson12}. Fouesneau et al.~(2014) derived the ages and masses of the clusters in the survey and found a flat \dndt\ distribution for ages younger than $\sim$70 Myr, again indicating the small role of disruption in setting the shape of the age distribution in this galaxy.

In the present work we study the age distributions of clusters in the nearby \citep[$\sim4.5$~Mpc,][]{thim03}, face-on spiral galaxy M83, based on seven pointings observed with the Wide Field Camera 3 (WFC3) onboard the Hubble Space Telescope (HST).  This study aims to complement and extend the previous studies of \citet[][hereafter C10]{chandar10}, \citet[][hereafter F12]{fouesneau12} and \citet[][hereafter B11 and B12, respectively]{bastian11,bastian12}.  C10 reported a steep age distribution of clusters and associations in the first of the seven fields analyzed here (F1, see Fig.~1), with $\dndt \sim t^{-0.8}$.  B12 analyzed the same region (F1) and came to similar conclusions. However, they included another pointing further away from the galaxy centre (F2), and found $\dndt \sim t^{-0.5}$.  Here, we extend the B11 and B12 analysis by including five further WFC3 pointings of M83. For this study, we try to avoid the problem introduced by a luminosity-limited sample and we will only compare age distributions of mass-limited samples. 

This paper is organized as follows.  In \S~\ref{sec:obs} we present the data and techniques used and in \S~\ref{sec:results} we show our main results.  We discuss the implications of our findings in \S~\ref{sec:discussion}.

\begin{figure}
\includegraphics[width=8.5cm]{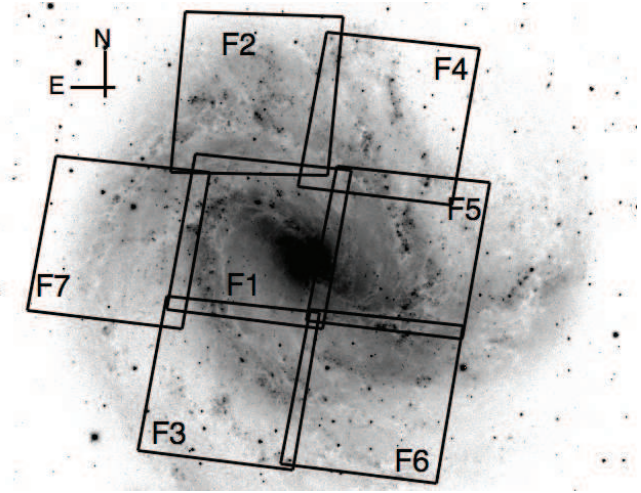}
\caption{M83 R-band image \citep{meurer06} with overlapping lines representing the areas covered with the HST. Taken from NASA/IPAC Extragalactic Database (NED).}
\label{fig:mosaic}
\end{figure}

\section{Observations and Techniques}
\label{sec:obs}

\subsection{HST WFC3 Imaging}
\label{sec:hst}

For the present work we use archival HST/WFC3 images from programme IDs 11360 (PI O'Connell) and 12513 (PI. Blair).  The dataset consists of imaging with the F336W, F438W, F547M, F657N, and F814W filters.  We will refer to these filters as U, B, V, H$\alpha$, and I, respectively, although no transformations were applied.  For one of the fields (Field~1) the F555W filter was used instead of the F547M filter.  Further details on the data used here will be given in a future work, Silva-Villa et al.~(in preparation).  

\subsection{Cluster Selection and Analysis}
\label{sec:selection}

In order to develop a catalogue of stellar clusters and groups, we apply the same methodology as used in B11 and B12.  First, we ran Source Extractor \citep{bertin96} over the V-band images for each field to obtain an initial catalogue of sources. We then carried out aperture photometry on this catalogue using apertures of 1 and 3 pixels ($\sim 0.9$ and $2.7$ pc), in order to measure the concentration index (CI - the magnitude within 1 pixel minus the magnitude in the 3 pixel aperture) of the sources (e.g., C10).  The resulting histogram of the CI values showed a clear peak at $CI=1.05$ mag, with a dispersion of 0.07 mag for all fields, and a tail of sources with larger CI.  The peak consists of unresolved sources (foreground stars as well as stellar sources in M83) while the tail consists of resolved sources.  We include only sources with CI indices above 1.25 mag, which, from tests carried out with ISHAPE \citep[][]{larsen99}, corresponds to a resolved source with an effective radius of $\sim$0.5~pc \citep[e.g.][]{konstantopoulos13}. 

We visually inspected each resolved source to classify it as a bonafide cluster (centrally concentrated, smooth flux distribution, no close neighbours), a likely group/association or blended source (multiple centres, non-centrally concentrated, severely crowded), or as a bad source (saturated star, too close to an image edge).  We will refer to these groups as class 1, 2 and 3, respectively.  See B12 for examples of class 1s (clusters) and class 2s (potential clusters, stellar groups or associations or asterisms). The results we present here are based on class 1 and 2 objects. Additionally, we removed the inner $450$~pc of the galaxy (as in B11,B12) given that the detection limit there is significantly worse than in the rest of the galaxy, resulting in a strong bias towards only detecting young ($<10$~Myr) clusters (see Fig.~8 of F12).

Aperture photometry was performed on the final catalogue with an aperture size of 5 pixels ($\sim4.4$ pc) and an inner and outer background radius of 8 to 10 pixels.  Zeropoints were taken from the STScI website\footnote{$http://www.stsci.edu/hst/wfc3/phot\_zp\_lbn$}.  The observations spanned $\sim3$~years and were processed with slightly different reduction pipelines.  In order to make sure that the photometric system was consistent between all the images, we compared various U, B, V, I, and H$\alpha$ sources ($\sim30$) in the overlap regions between the fields.  We found small systematic offsets between Fields 1 and 2 compared to the rest of the fields in the U and V-band magnitudes, and so we applied small corrections ($<0.05$~mag) to the photometry in these bands for these fields.

We then compared the photometry of each cluster to the \citet{zackrisson11} simple stellar population (SSP) models with solar metallicity, and a \citet{kroupa01} stellar initial mass function, using the $\chi^2$ minimization routine presented in \citet{adamo10} (i.e., the same models and technique used in B11 and B12).

\section{Results}
\label{sec:results}

For the present study we restrict our analysis to clusters with derived masses above $5000$~\msun, in order to minimize the effects of stochastic sampling \citep[see e.g.][and references therein]{barbaro77,popescu09,fouesneau10,silvavilla11,anders13}. The application of this limit results in a mass limited cluster sample to ages of $\sim400$~Myr. We limit further our analysis to the most recent $\sim300$~Myr (Log($\tau$/yr) $\sim8.5$). A cluster with an observed mass of $\sim5000\msun$ (or higher) and an age of $\sim$300 Myr (or younger) has a V-band of $\sim-5$ mags (or brighter). The imposed lower mass limit, reduces the affect of stochasticity in the fitting procedure \citep[e.g.][]{fouesneau10} and ensures that we are mass limited (i.e., not luminosity limited)

\begin{figure}
\includegraphics[width=8.5cm]{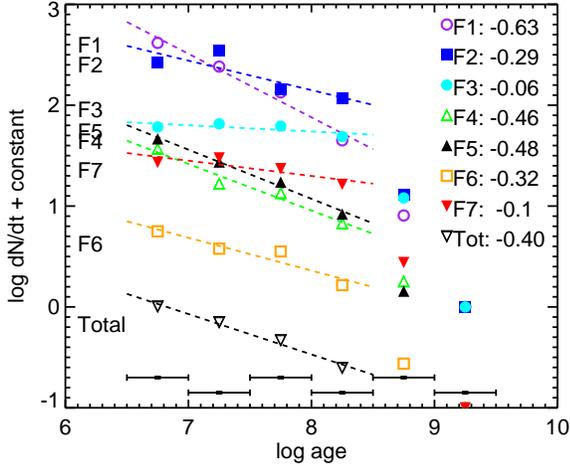}
\caption{The age distribution (\dndt) of clusters (classes 1 and 2) in each field with masses above 5000 Msun. Bins with widths of 0.5 dex were used.  The same bins were used for all fields for direct comparison.  The dashed lines represent fits to the data (carried out over the region shown). Top right are the slope of each \dndt\ distribution. }
\label{fig:dndt1}
\end{figure}

Figure~\ref{fig:dndt1} shows the resulting age distributions of the clusters for each field, where we consider both class 1 and class 2 sources.  We used bin widths of 0.5 dex. We note that the values do not change drastically if we use only class 1 objects. Assuming a simple power-law of the form $\dndt \sim t^{\zeta}$ \citep[e.g.][]{whitmore07}, we estimated the slope of the age distributions and present out results in the figure. We find that the fields with the steepest slope are Fields~1, 4 and 5, with $\zeta=-0.6$, $-0.5$ and $-0.5$, respectively. In contrast, both Fields 3 and 7 have $\zeta\sim-0.1$, showing the large difference between the fields.  

\begin{figure}
\includegraphics[width=8.5cm]{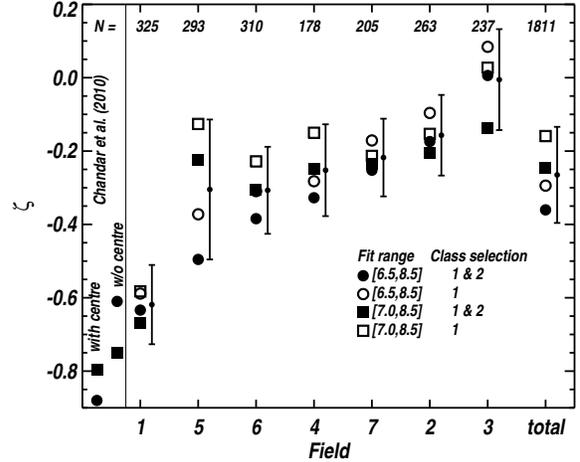}
\caption{The indices from maximum likelihood fits to each field of the form $\dndt \sim t^{\zeta}$.  The symbols represent different choices of the classes of sources used in the fits and different age ranges where the fit is made. Note that the fields are ordered in increasing $\zeta$ values.  Numbers (N) at the top are the number of clusters for the filled circles in the legend.  ``Total" represents the full catalogue.  Additionally, we plot the results from the C10 catalogue (with symbols that represent both class 1 and class 2 sources). The error is a combination of the standard deviation of the scatter and the random error in the fit.}
\label{fig:dndt2}
\end{figure}

In order to provide a more statistically and robust analysis that avoids the pitfalls inherent in binning discrete data, we also applied a maximum likelihood fit to the distributions. Different age ranges and classes (i.e. class 1 only, or class 1 and 2 combined) were used in the fit. The results are shown in Fig.~\ref{fig:dndt2}, where the recovered \dndt\ slopes are depicted as a function of increasing $\zeta$ value. We conclude from Figs.~\ref{fig:mosaic} and \ref{fig:dndt2}  that the shallower values are recovered in the outermost fields. We obtain a significantly broad range of $\zeta$ values in Field 3 and 5 depending on the cluster selection. This dispersion is likely due to a variation in the number of objects at young ages ($log(\tau/yr)\le 7.0$), which shifts the first bin value up/down, changing the value of $\zeta$. The dispersion suggests that the \dndt\ distribution is not well approximated by a single power-law distribution. The estimated error of $\zeta$ for the maximum likelihood method is the sum in quadrature of the Poissonian error (typically 0.1, based on monte-carlo simulations of cluster samples with the same number of objects) in the fit due to the finite number of objects, and the standard deviation in the scatter for selection criteria and fitting range ($\sim0.04 \ - \sim0.16$).

\subsection{Comparison with previous results}
\label{sec:comparison}

Field 1 was studied in detail in C10, F12 and B12 so we can compare our distributions to those.  C10, F12 and B12 made their catalogues public so that a direct comparison can be made, although here we use only C10 for comparisons.  In Fig.~\ref{fig:dndt2} we show the resulting maximum likelihood fits to the C10 catalogue, where we have removed the inner $450$~pc of the disk, to be consistent with the catalogue presented here.  In this particular case (i.e. C10 catalogue, removing the inner centre of the galaxy, where the two samples overlap), our results show excellent agreement with C10, showing that the age distribution is not heavily affected by source selection or analysis techniques.  This gives us confidence to expand out into other regions, applying the same methods and techniques in order to see if we find the same results.

\citet{silvavilla11} analysed WFPC2/ACS HST imaging of two fields within M83 (one field overlapping with Fields 3 and 6, and  another overlapping with Fields 1, 2, 4 and 5).  They find $\zeta = -0.26$, in excellent agreement with our estimation for the total sample (however using a different age range).  

F12 analyzed Field~1, using the catalogue of C10 but with ages, masses and extinctions estimated using stochastically sampled SSP models.  They find $\zeta=-1.0$, however they included the inner 450~pc which, due to the worse detection limit, is strongly biased towards detecting young clusters (see Fig.~8 in F12).

Comparison of Fields 1 and 2 with B12 shows that the current age distributions are shallower by $\Delta(\zeta) \sim 0.2$ than previously reported.  Upon further tests we found a bug in the extinction estimation routine used in B12. However, since the bulk of the cluster population has very low extinction $E(B-V)\sim0.2$ mag, the impact on the derived analysis is extremely low (i.e. $\Delta(E)\sim0.2$).  Running the previous dataset through the corrected routine results in age distributions for both Fields 1 and 2 that agree closely with those presented here.  The coding bug and its implications on the results of B12 will be discussed in an upcoming paper, Silva-Villa et al.~(in preparation).

\section{Discussion and Conclusions}
\label{sec:discussion}

As discussed in \S~1, the age distribution of clusters is a convolution of the cluster formation history and cluster disruption. \citet{fall05,whitmore07,chandar10,fall12}, among other authors, have suggested a Universal (or quasi-Universal) age distribution in a form of a steep power-law with an index $\zeta$ between -0.8 and -1.0.  These authors found this in a variety of galaxies, therefore they conclude that this rapid disruption must be independent of the environment. Since it is unlikely that all galaxies would have undergone a rapid increase in the cluster formation rate over the past few hundred Myr, they argue that the steepness of the distributions is caused by rapid cluster disruption. However, if true, there are important discrepancies between this environmental independence and the results from analytic models based on N-body simulations \citep[e.g.][]{baumgardt03,gieles06}, and it will be require a new framework to understand the evolution of star clusters.

It is worth noting that the Universal model was originally developed to explain the stellar cluster population of the merging, starburst galaxy, the Antennae, where it was assumed that the cluster formation rate and history was constant over the past few hundred Myr, contrary to expectations of numerical simulations \citep[e.g.][]{baumgardt03,bastian09,chien10,kruijssen11}.

We have attempted to reduce systematics between the fields, by adopting $(i)$ the same photometric procedure, $(ii)$ the same SED models, and $(iii)$ fitting procedure using identical age bins. Additionally, we have also carried out a maximum likelihood fit to the distributions to avoid binning altogether. Our results from these two different methods are in excellent agreement.  

The results found for the seven fields in M83 do not agree with the predictions of the Universal model, as we find significantly shallower age distributions, with values of $\zeta$ between $-0.1$ and $-0.61$. These variations indicate a strong correlation with the local ambient environment.  These shallower distributions agree with analyses of the SMC \citep{gieles07,degrijs08,portegieszwart10}, LMC \citep{baumgardt13,degrijs13}, M51 \citep{gieles05}, and NGC~2997 (Ryon et al. in prep.) cluster populations, as well as previous M83 results \citep[B11; B12;][]{silvavilla11}.

Is there a way in which the observations presented here could be made to fit with the Universal scenario?  One could argue that the star formation history in M83 has been decreasing over the past 200-300~Myr, which would counteract the observational effects of cluster disruption.  In order to change the observed \dndt\ distributions to be consistent with the Universal expectations, the star formation rates would have had to have decreased by a factor of 2-3 (Field~1) to 10 (Fields 3 and 7) over the past 100~Myr. However, based on the results presented by \citet{silvavilla11} and B12, based on resolved stellar populations, such changes are unlikely.

Under the assumption that the cluster formation history/rate has not dramatically changed in the time interval we are studying here \citep[based on resolved stellar population studies, e.g.][]{silvavilla11}, we conclude that the cluster population of M83 has not been dramatically affected by cluster disruption, in contrast to expectations from the Universal scenario, where clusters are disrupted at a high rate independent of their ambient environment \citep[e.g.][]{chandar06,chandar10,fall12}. 

On the other hand, the mass and environmentally dependent disruption model \citep[MDD model,][]{lamers05} predicts that the slope of the age distribution will be strongly tied to the tidal fields and the surface density of giant molecular clouds (GMCs), leading to a strong correlation with the ambient environment. Under the assumption of a (roughly) constant cluster formation history, if the tidal field is strong and the GMC density is high, cluster disruption is expected to be stronger, leading to steeper age distributions. On the contrary, if the tidal field is weak and the GMC density is low, there will be less disruption, and the age distribution will be flatter. We find that the inner fields, where the tidal forces and GMC density are highest \citep[c.f.][]{bastian12} have steeper age distributions, while in the outer regions they are flatter, in excellent agreement with predictions from the models \citep[e.g.][]{lamers05,kruijssen11}.   We will further quantify the relation between environment and disruption using the M83 cluster dataset in a future study.

A study of the resolved stellar population, along with the presentation of the full catalogue and a further study of cluster disruption, will be presented in a future work, Silva-Villa et al. (in preparation).  Additionally, we will study the relation between star and cluster formation, and the dependence of environment in Adamo et al. (in preparation).

\section*{Acknowledgments}

ES-V is a postdoctoral fellow supported by the Centre de Recherche en Astrophysique du Qu\'ebec (CRAQ). NB is partially funded by a Royal Society University Research Fellowship. EZ acknowledges funding from the Swedish National Space Board and the Swedish Research Council. ES-V, NB and MF acknowledge the hospitality of the Aspen Center for Physics, which is supported by the National Science Foundation Grant No. PHY-1066293.

\end{document}